\documentclass[aps,physrev,twocolumn,superscriptaddress]{revtex4-2}

\usepackage{amsmath}
\usepackage{amssymb}
\usepackage{amsthm}
\usepackage{amsfonts}
\usepackage{listings}
\usepackage{physics}
\usepackage{enumerate}
\usepackage{latexsym}
\usepackage{psfrag}
\usepackage{bm}
\usepackage{graphicx}
\usepackage[caption=false]{subfig}
\usepackage{blkarray}
\usepackage{array}
\usepackage{color}
\usepackage[normalem]{ulem}
\usepackage{hyperref}
\usepackage{cleveref}
\usepackage{dsfont}
\usepackage{tabularx}
\usepackage{multirow}
\usepackage{xfp}
\usepackage{calc}
\usepackage[nomessages]{fp}
\usepackage{booktabs}
\usepackage{siunitx}  

\newcommand{\tempone}{\SI{25}{\celsius}}
\newcommand{\temptwo}{\SI{100}{\celsius}}
\newcommand{\tempthree}{\SI{300}{\celsius}}
\newcommand{\tempfour}{\SI{500}{\celsius}}

\graphicspath{ {./figures/} }

\begin{document}


\title{Connecting Microseismicity to Lithology via a Model of Slip Avalanches}


\author{Ethan A. Mullen}
\affiliation{Department of Physics and Anthony J. Leggett Institute for Condensed Matter Theory, University of Illinois at Urbana-Champaign, 1110 West Green Street, Urbana, Illinois 61801, USA}
\author{Yanbo Wang}
\affiliation{Integrated Multi-Physics Lab, Department of Mechanical Engineering, The University of Utah, Salt Lake City, UT 84112, USA}
\author{Jordan J. Sickle}
\affiliation{Department of Physics and Anthony J. Leggett Institute for Condensed Matter Theory, University of Illinois at Urbana-Champaign, 1110 West Green Street, Urbana, Illinois 61801, USA}
\author{Frank W. DelRio}
\affiliation{Material, Physical, and Chemical Sciences Center, Sandia National Laboratories, Albuquerque, NM 87123, USA}
\author{Pania Newell}
\affiliation{Integrated Multi-Physics Lab, Department of Mechanical Engineering, The University of Utah, Salt Lake City, UT 84112, USA}
\author{Anastasia G. Ilgen}
\email[]{agilgen@sandia.gov}
\affiliation{Geothermal Research Department, Sandia National Laboratories, Albuquerque, NM 87123, USA}
\author{Karin A. Dahmen}
\email[]{dahmen@illinois.edu}
\affiliation{Department of Physics and Anthony J. Leggett Institute for Condensed Matter Theory, University of Illinois at Urbana-Champaign, 1110 West Green Street, Urbana, Illinois 61801, USA}

\date{\today}

\begin{abstract}
Fluid injection into the earth’s crust can induce small and frequent earthquakes in the subsurface. Predicting their sizes and temporal occurrences via statistical analysis is crucial for safe operations in unconventional oil and gas recovery, enhanced geothermal systems, and geologic carbon storage. Here we show that the slip statistics observed over drastically different spatial scales, namely m-scale microseismic observations and nm- to $\mu$m-scale nanoindentation experiments, are consistent with the predictions of a simple micromechanical model of slip avalanches in slowly deforming solids. Microseismic catalogs extracted from high-pressure fluid injection operations into geological basins with various lithologies and nanoindentation experiments on shale across a wide range of temperatures and mineral compositions yield statistics consistent with model predictions. This universality across materials, temperatures, and scales is consistent with the prediction that the slip statistics result from only a few basic properties. Previously debated deviations of the statistics in layered sedimentary formations are explained by finite-size and stress-integrative effects resulting from mechanically weak bedding planes. The slip statistics therefore provide important information about the structure and scales of the bedding planes. Conversely, the basin structure can also be used to predict the probability distribution for the sizes of triggered microseismic events.
\end{abstract}

\maketitle

\section{\label{sec:intro}Introduction}
Significant technological progress has recently been made in deep-well drilling and subsequent fluid injection for hydraulic fracturing (HF) in unconventional oil and gas recovery, hydraulic shearing or hydroshearing (HS) in enhanced geothermal systems (EGS), and carbon dioxide injection in geologic carbon storage (GCS)\cite{2022_Ma_et_al, 2016_Olasolo_et_al, 2022_Temizel_et_al}. These applications inject fluids into the earth at pressures high enough to stimulate fracture growth and/or shear slip in the subsurface. Both during and after the injections induced earthquakes are observed with moment magnitudes mostly below three \cite{2018_Eaton, 2020_Schultz_et_al, 2019_Li_et_al}, though quakes of moment magnitude greater than four have also been reported. These small, frequent earthquakes occur close to the wellbore (typically within a few hundred meters \cite{2019_Li_et_al}), serving as sensitive subsurface signals of local stress state changes due to fluid injection or withdrawal. As such, microseisms are crucial tools for examining fracture growth and ensuring that fracture networks that can result in geohazards are not opened \cite{2018_Eaton}.

High-resolution measurement techniques have enabled the collection of large microseismic magnitude catalogs during and after fluid injection operations. Over a broad range of sizes, the size distributions of these moment magnitudes are described by a power law (similar to the Gutenberg-Richter relationship for larger earthquakes) \cite{1944_Gutenber_Richter}. Namely, 
\begin{equation}
\label{eq:GR_equation}
\log(N)=a-bM,
\end{equation}
where $M$ is the dimensionless moment magnitude of an earthquake (which grows with the logarithm of the total slip in an earthquake, see below), $N$ is the number of events in a catalog with magnitudes $\ge M$, and $a$ and $b$ are dimensionless numbers that quantify the overall rate of seismicity in a catalog and the relative number of large to small events, respectively. Given that the $b$-value describes the relative likelihood of large and therefore potentially hazardous induced quakes, it encodes crucial information for hazard and efficiency assessments.

For global earthquake catalogues with $M\gtrsim 5.60$ \cite{2004_Bird_Kagan} and natural fault systems with $4.00 \le M \le 8.00$ \cite{1993_Frohlich_Davis, 2010_Kagan}, one finds $b\approx1.00 \pm 0.30$. However, this value is thought to depend on several factors, including the fault regime and the state of stress and its temporal variations \cite{2018_Eaton}. Schorlemmer \emph{et al.} \cite{2005_Schorlemmer_et_al} reported that the $b$-value varies systematically with faulting style for large quakes: thrust faults exhibit the lowest $b$-values $(b \approx 0.70)$ and normal faults exhibit the highest $(b \approx 1.10)$, with strike-slip faults exhibiting values in between $(b\approx 0.90)$. Later, this variation was explained by a stress-dependent cutoff to the power-law distribution of earthquake moments \cite{2015_Uhl_et_al}. 

In contrast, $b$-values extracted from microseismic catalogs vary widely across different target formations. HF and EGS catalogs have yielded $b$-values anywhere in the range 0.00 -- 4.00 \cite{2010_Downie_et_al, 2011_Maxwell_et_al, 2018_Maghsoudi_et_al, 2016_Maghsoudi_et_al, 2019_Zorn_WV_marcellus_shale, 2017_Zorn_MSEEL, 2017_Zorn_Pennsylvania}, with $b\approx1.00$ for EGS catalogs and $b\approx2.00$ for HF catalogs. This difference may reflect the distinct rock types involved in these operations. HF operations are typically conducted in layered sedimentary formations containing relatively weak mechanical bedding planes that limit shear failure at their interfaces \cite{2003_Miskimins_Barree}, whereas EGS projects focus on more homogeneous, mechanically strong basement rocks that lack such regular layering \cite{2008_Huh_et_al}.

The variation in $b$-value is also spatiotemporal in nature within a particular lithological setting. For instance, different $b$-values may be measured at different locations around the same wellbore \cite{2011_Maxwell_et_al, 2019_Zorn_WV_marcellus_shale, 2017_Zorn_MSEEL, 2012_Hurd_Zoback}, and variation is also seen during and after fluid injection \cite{2010_Downie_et_al}. Moreover, some authors find that $b$-values depend on rock mechanical properties \cite{2019_Zorn_WV_marcellus_shale, 2017_Zorn_MSEEL, 2017_Zorn_Pennsylvania} and even observation well locations \cite{2016_Grob_vanderBaan}. While geophysical characteristics likely lead to some of the observed $b$-value variability, the statistical approaches used in the literature to extract these values may also significantly contribute to the $b$-value spread. Further discussion of these approaches and their effects can be found in the Discussion and Methods sections. Here we use modeling, experiments, and observation to conjecture a rigorous explanation for the physical factors that determine $b$-values.

The goal of this article is to demonstrate that the observed lithological dependence of the $b$-value extracted from microseismic catalogs is consistent with the predictions of a simple, nonparametric mean-field model of slip avalanches \cite{1998_Dahmen_et_al, 2009_Dahmen_et_al} in combination with geophysical observations, nanoindentation experiments, and statistical considerations. 

This model assumes that any sheared solid has weak spots (i.e., dislocations in crystals or shear transformation zones (STZs) in amorphous solids). Each weak spot remains in place until the local stress exceeds a local slip threshold. When a weak spot slips on a glide plane, in a shear band, or on an earthquake fault, it relaxes and distributes the local stress to the remaining weak spots (or planes in a layered system). It can thereby trigger other weak spots to slip, resulting in a slip avalanche, observable as an earthquake or a microseismic event in the present context \cite{2015_Uhl_et_al, 1998_Dahmen_et_al, 2009_Dahmen_et_al}.  In the most general case, this model incorporates a single parameter $\epsilon$ (known as weakening) to encode information about the dynamical behavior of slip sites: after a site $i$ slips, its failure threshold is reduced by an amount proportional to $\epsilon_i$. In the present case, we have assumed that the weakening is negligible. Using renormalization group tools one can show that the mean-field approximation (where all weak spots are coupled equally to each other) yields the same scaling behavior for the slip statistics as those obtained from full 3D elasticity \cite{2009_Dahmen_et_al, 1997_Fisher_et_al}. In the mean-field limit, the model describes systems where the slip is localized in slip planes. In this case many details of the interactions (such as quadrupolar characteristics) become irrelevant for the slip statistics so that the simplifying assumptions of this model are justified \cite{2009_Dahmen_et_al, 1997_Fisher_et_al}.

Additionally, the model predicts several scaling laws \cite{2009_Dahmen_et_al, 2013_Leblanc_et_al}, the most relevant of which being that of the complementary cumulative distribution function (CCDF) of event sizes (see Results section for exact definitions of event sizes in both the nanoindentation and microseismic contexts)
\begin{equation}
\label{eq:ccdf_definition}
C_S(s)\sim s^{-(\kappa-1)} G(s/s_{\text{max}}),
\end{equation}
where $C_S(s)$ represents the probability that the slip avalanche size $S$ takes on a value greater than or equal to $s$, $\kappa$ is a critical exponent, $G(x)$ is the universal scaling function 
\begin{equation}
\label{eq:scaling_function}
G(x)\sim x^{\kappa-1}\int_x^\infty e^{-At}t^{-\kappa}dt,
\end{equation}
which imposes a cutoff on the distribution $C_S(s)$ and contains a material-dependent constant $A$\cite{2015_Uhl_et_al}, and $s_{\text{max}}$ can depend on system features such as stress, pressure, temperature, fault size, etc. The critical exponent $\kappa$ is a universal quantity expected to hold across all systems described by the theory regardless of experiment-specific scales and/or parameters \cite{2009_Dahmen_et_al, 2001_Sethna_et_al}. The probability density function (PDF), used for all fitting in this work, is related to the CCDF via $P_S(s) = \frac{d}{ds}\left(1-C_S(s)\right)$.

This model has seen great success in explaining the observations of avalanche-like events present when materials across a vast range of physical scales are slowly deformed. Nanocrystals, bulk metallic glasses (BMGs), granular materials, lamellar solids, rocks, and even the earth itself have all demonstrated slip statistics consistent with model predictions \cite{2015_Uhl_et_al, 2022_Salners_et_al, 2024_Sickle_et_al_Mica, 2025_Sickle_et_al_BMGs}, and recent detailed work on mica and BMGs \cite{2024_Sickle_et_al_Mica, 2025_Sickle_et_al_BMGs} has demonstrated the applicability of the model to nanoindentation experiments. In the following we use lab-scale nanoindentation tests on shale samples from the Cane Creek formation in Utah to demonstrate the robustness of the theory across many materials and many orders of magnitude in size.

We begin by showing that nanoindentation experiments on shale samples extracted from the Cane Creek formation in Utah reveal failure statistics consistent with the simple mean-field theory (MFT) predictions. These experimental results demonstrate the existence of striking similarity between the shale samples’ pop-in distributions, confirming the theoretical prediction that system-specific details are irrelevant to the event statistics, and robustness to temperature up to \tempfour. We then bridge the scales by showing that the $b$-values obtained from microseismic catalogs spanning a range of lithological and operational settings are also consistent with MFT predictions. We propose that the mechanically weak bedding planes present in sedimentary formations give rise to finite-size and stress-integration effects through the distribution of layer thicknesses and the orientation-dependent stresses acting on slip planes, respectively (see Supplemental Information). These effects account for the elevated $b$-value of approximately $2.00$ observed in such formations. These results suggest that microseismic event statistics can, in turn, provide important information about the underlying lithology.

\section{\label{sec:results}Results}

\subsection{\label{subsec:nanoindentation}Comparing Nanoindentation Experiments to Model Predictions: Pop-in Size Distributions at Different Temperatures}
\begin{figure*}
\includegraphics[]{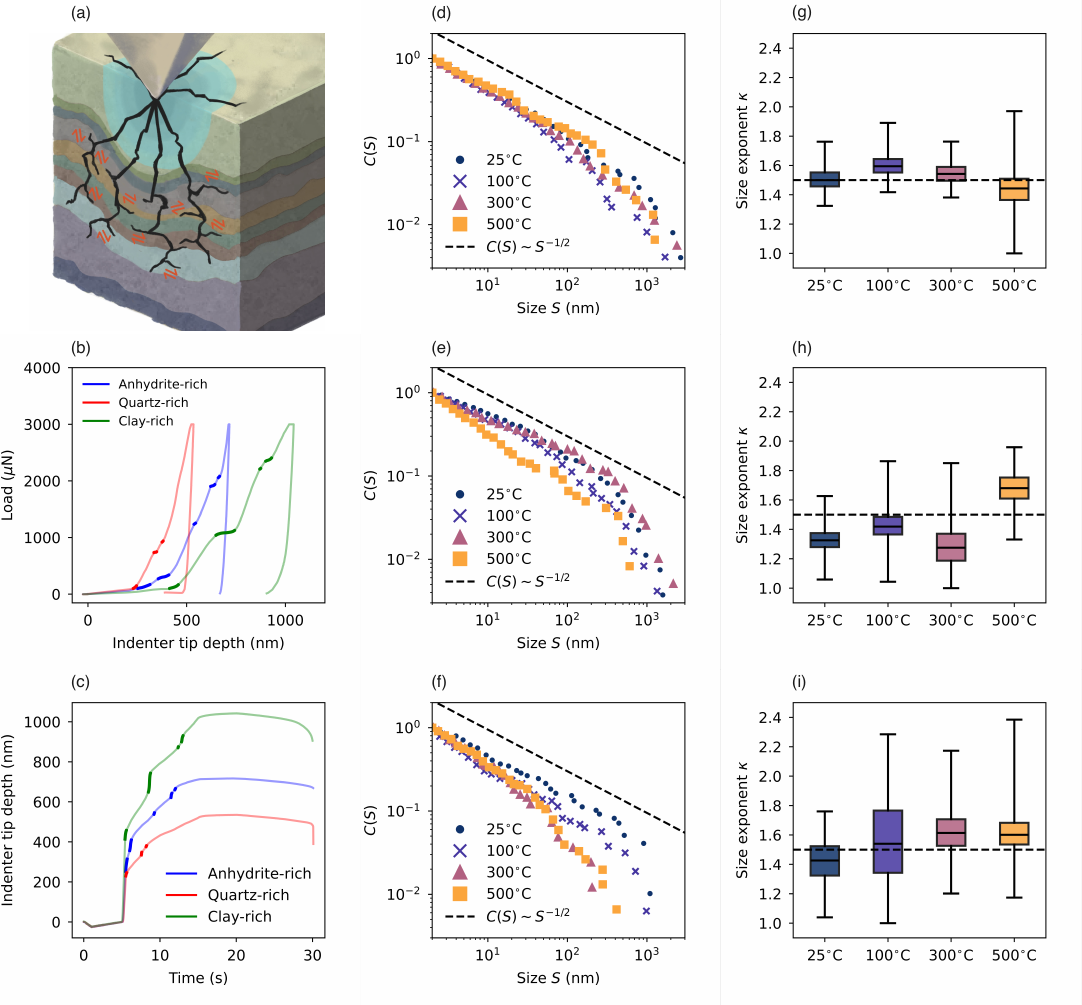}
\caption[]{
Nanoindentation of Cane Creek Shale demonstrates consistency with the mean-field model.
(a) Schematic of nanoindentation. A Berkovich tip is driven at constant loading rate perpendicular to the surface of a polished shale sample. 
(b) Load versus tip depth curves for all three shale samples. Highlighted are pop-in events greater than 2 nm.
(c) Tip depth versus time curves for all three shale samples. Highlighted are the same pop-in events shown in panel (b). 
(d)-(f) $C(S)$ for the shale samples ((d) anhydrite-rich, (e) clay-rich, (f) quartz-rich) across all four temperatures. The mean-field prediction $C(S) \sim S^{-(\kappa-1)}$ is shown as a reference slope on each panel. 
(g)-(i) Box-and-whisker plots showing fitted exponents for the shale samples ((g) anhydrite-rich, (h) clay-rich, (i) quartz-rich). Boxes indicate inner 50\% of fitted values while whiskers indicate inner 95\% of fitted values. See Methods for a description of the fitting method used to extract these exponents and their 95\% confidence intervals. 
}
\label{fig:figure_1}
\end{figure*}

Nanoindentation experiments were performed on three Cane Creek shale samples, each of unique mineral composition. Table~\ref{tab:mineral_composition} gives the mineral compositions of each sample; the mineral with largest mass contribution to each sample serves as its label. Experimental details are given in the Methods section. Example load versus displacement and displacement versus time traces taken from these indents are shown in Fig.~\hyperref[fig:figure_1]{\ref*{fig:figure_1}(b)} and Fig.~\hyperref[fig:figure_1]{\ref*{fig:figure_1}(c)}, respectively, for each of the three samples examined here. In both Fig.~\hyperref[fig:figure_1]{\ref*{fig:figure_1}(b)} and Fig.~\hyperref[fig:figure_1]{\ref*{fig:figure_1}(c)}, vertical tip displacement events (“pop-ins”) greater than a noise threshold of 2 nm are shown; see Methods and Supplemental Information for more details on this threshold. A pop-in is defined as a tip displacement event where the slope of the displacement versus time trace exceeds the median value in a sliding window around each point. (Other studies have identified pop-in events during the creep-like motion common in shale nanoindentation experiments \cite{2019_Shi_et_al, 2022_Wang_et_al, 2023_Charlton_et_al, 2024_Gu_et_al}.) 

\begin{table*}[t]
    \caption{Nanoindented shale mineral compositions. Mineral percentages (by mass) of each shale sample used in this work.}
    \label{tab:mineral_composition}
    \begin{tabular}{lccc}
        \toprule
        Mineral & \shortstack{Anhydrite-rich \\ (Depth: 9654.3 ft)} & \shortstack{Quartz-rich \\ (Depth: 9673.6 ft)} & \shortstack{Clay-rich \\ (Depth: 9720.5 ft)} \\
        \midrule
        Anhydrite  & 57\% & 0\% & 2\% \\
        Quartz  & 18\% & 56\% & 23\% \\
        Illite  & 4\% & $<1\%$ & 0\% \\
        Interlayered illite-smectite  & 0\% & 0\% & 26\% \\
        Calcite  & 0\% & 26\% & 23\% \\
        Dolomite  & 15\% & 15\% & 2\% \\
        Halite  & $<1\%$ & 0\% & 3\% \\
        Plagioclase  & 2\% & $<1\%$ & 5\% \\
        K (potassium)-feldspar & 2\% & 2\% & 7\% \\
        Chlorite & $<1\%$ & 0\% & 9\% \\
        \bottomrule
    \end{tabular}
\end{table*}

\begin{table*}[t]
    \caption{Indent and pop-in counts. For each sample and at each temperature the number of indents used for analysis is shown in the first column division and the number of detected pop-ins is shown in the second column division.}
    \label{tab:indent_popin_counts}
    \centering
    \begin{tabular}{lcccccccc}
        \toprule
        & \multicolumn{2}{c}{\tempone}
        & \multicolumn{2}{c}{\temptwo}
        & \multicolumn{2}{c}{\tempthree}
        & \multicolumn{2}{c}{\tempfour} \\
        \cmidrule(lr){2-3}
        \cmidrule(lr){4-5}
        \cmidrule(lr){6-7}
        \cmidrule(lr){8-9}
        Sample
        & Indents & Pop-ins
        & Indents & Pop-ins
        & Indents & Pop-ins
        & Indents & Pop-ins \\
        \midrule
        Anhydrite-rich & 97 & 250 & 100 & 246 & 80 & 179 & 121 & 152 \\
        Clay-rich & 99 & 268 & 100 & 241 & 100 & 197 & 100 & 121 \\
        Quartz-rich & 100 & 98 & 100 & 159 & 100 & 83 & 102 & 152 \\
        \bottomrule
    \end{tabular}
\end{table*}

To measure thermal effects, each sample was indented at four temperatures ranging from \tempone--\tempfour. The total number of indents and the pop-in count for each sample at each temperature can be found in Table~\ref{tab:indent_popin_counts}. Figs.~\hyperref[fig:figure_1]{\ref*{fig:figure_1}(d)}-\hyperref[fig:figure_1]{\ref*{fig:figure_1}(f)} show the pop-in size CCDFs for each sample across all four temperatures. For pop-ins greater than both 2 nm in size and 50 ms in duration, power-law scaling in agreement with the predictions of the mean-field model is evident: the model predicts universal exponent values of $\kappa=1.50$ and $\alpha=2.00$ for the power-law size and duration CCDFs $C(S)\sim S^{-(\kappa-1)}$ and $C(T)\sim T^{-(\alpha-1)}$, respectively, which fall well within the corresponding $95\%$ confidence intervals of the fitted exponents. We have chosen to only show the size distributions here for direct comparison with the microseismic data; duration distributions and their best-fit exponents can be found in the Supplemental Information. Figs.~\hyperref[fig:figure_1]{\ref*{fig:figure_1}(d)}-\hyperref[fig:figure_1]{\ref*{fig:figure_1}(f)} demonstrate that all distributions exhibit power-law scaling for approximately two orders of magnitude, and Figs.~\hyperref[fig:figure_1]{\ref*{fig:figure_1}(g)}-\hyperref[fig:figure_1]{\ref*{fig:figure_1}(i)} each show the spread of fitted exponents for their respective pop-in collections. More information on the distribution fitting technique used to extract these exponents and their confidence intervals can be found in the Methods section.  

\subsection{\label{subsec:microseismic}Comparing Microseismic Activity to Model Predictions}
We analyzed seven microseismic data sets, two of which were taken from sedimentary formations and five from non-sedimentary formations (descriptions of each data set can be found in the Supplemental Information). The latter formations are generally composed of mechanically strong, brittle rocks such as volcanic lava flows, crystalline metamorphic rocks, welded tuffs, and crystalline granites. In contrast to sedimentary rocks such as shales, these rocks do not possess extensive layering. The CCDFs for these microseismic catalogs can be found in Fig.~\hyperref[fig:figure_2]{\ref*{fig:figure_2}(a)} and Fig.~\hyperref[fig:figure_2]{\ref*{fig:figure_2}(b)}. Four of the five non-sedimentary datasets come from EGS projects across the United States, while the fifth dataset comes from a carbon dioxide injection project targeting the Mt. Simon Sandstone in Decatur, Illinois. Due to different magnitude types reported across these datasets (see descriptions below), the moment magnitudes $M$ were first converted to effective scalar seismic moments $M_0^*$ using Eq. (\ref{eq:moment_and_mag}) \cite{2024_Pardo-Iguzquiza_Dowd, 2019_Zoback_Kohli} for ease of comparison with model predictions: 
\begin{equation}
\label{eq:moment_and_mag}
M=\frac{2}{3}\left(\log_{10}M_0^*-9.05\right).
\end{equation}
These effective scalar seismic moments were then rescaled by their respective maxima $M_{0,\text{max}}^*$ to aid visualization.

While there are many different seismic magnitude scales (including local magnitudes $M_{\text{L}}$, surface-wave magnitudes $M_{\text{S}}$, and body-wave magnitudes $m_{\text{b}}$), the scalar seismic moment $M_0$ is arguably the most definitive representation of earthquake size given its dependence on only absolute physical parameters independent of local scales. Seismic moments also allow for more direct comparison with both the theory and nanoindentation results, and we chose SI units for the conversion out of convenience. The standard definition of seismic moment is
\begin{equation}
\label{eq:seismic_moment}
M_0=\mu dA,
\end{equation}
where $\mu$ is the shear modulus of the rock, $d$ is the average slip distance, and $A$ is the slip area. More general definitions that depend on the slipping surface or the moment tensor can be found in Fisher \emph{et al.} \cite{1997_Fisher_et_al} and Silver and Jordan \cite{1982_Silver_Jordan}, respectively. For typical shale formations, for example, $\mu$ will be on the order of $2.2 \times 10^6$ psi $\left(\approx 15 ~ \text{GPa}\right)$ \cite{2012_Warpinski_et_al}. The fitted effective seismic moment probability distribution exponents $\kappa$ can be converted to $b$-values using Eq. (\ref{eq:b_to_kappa}) \cite{2010_Kagan, 2024_Pardo-Iguzquiza_Dowd}:
\begin{equation}
\label{eq:b_to_kappa}
b=\frac{3}{2}\left(\kappa-1\right).
\end{equation}
A description of the fitting method used to extract these exponents can be found in the Methods section. 

In Figs.~\hyperref[fig:figure_2]{\ref*{fig:figure_2}(a)}-\hyperref[fig:figure_2]{\ref*{fig:figure_2}(b)} we see remarkable consistency in scaling behavior between catalogs for each of the two formation types, and in Fig.~\hyperref[fig:figure_2]{\ref*{fig:figure_2}(a)} and Fig.~\hyperref[fig:figure_2]{\ref*{fig:figure_2}(c)} we see that the two catalogs from sedimentary formations exhibit nearly identical scaling behavior across a nearly identical range of seismic moments. These catalogs’ typical fitted exponent $\kappa^*=2.35$ is representative of the generally higher $b$-values noted in the literature as mentioned earlier. In Fig.~\hyperref[fig:figure_2]{\ref*{fig:figure_2}(b)} and Fig.~\hyperref[fig:figure_2]{\ref*{fig:figure_2}(d)}, we see similar scaling consistency. Note that we have rescaled each catalog’s seismic moments due to their different reported magnitude types. 

\section{\label{sec:discussion}Discussion}

\subsection{\label{subsec:nanoindentation}Nanoindentation}
Figs.~\hyperref[fig:figure_1]{\ref*{fig:figure_1}(d)}-\hyperref[fig:figure_1]{\ref*{fig:figure_1}(f)} show that the distributions of pop-in sizes obtained during nanoindentation experiments are consistent with MFT predictions. The CCDF was chosen for visualization purposes and all power-law slopes shown in Fig.~\hyperref[fig:figure_1]{\ref*{fig:figure_1}} are references to the best-fit exponents extracted from the maximum-likelihood approach described in Methods. All three samples exhibit nearly identical power-law scaling and maximum pop-in sizes of approximately $1 ~ \mu \text{m}$. As expected, no general trend with temperature is observed because the minerals present in these shale samples exhibit phase changes only at temperatures well above \tempfour ~ (see Supplemental Information). For each sample, however, there are few pop-ins across a broad range of sizes. This limited dataset of pop-in events leads to fitting challenges which result in a noteworthy spread in fitted exponents. Power-law fitting methods are known to be quite sensitive to minimum and maximum values used when fitting, especially when the fitted dataset is small \cite{2009_Clauset_et_al, 2013_Deluca_Corral}. Moreover, since the mean pop-in duration is expected to scale with size as $T\sim S^{1/2}$, the range over which power-law behavior is observed in durations is expected to span roughly half as many decades as in that of the sizes. Despite these limitations, the consistent scaling in pop-in sizes across samples and temperatures as well as the proximity of each fitted exponent to its value predicted by the model demonstrate consistency with the theory. Future work using an increased number of indents at higher sampling rate will allow for testing the existence of a power-law scaling relationship between pop-in sizes and durations as well as scaling collapses \cite{2025_Sickle_et_al_BMGs, 2018_Dahmen_Wright}, with the latter being much stricter tests of critical behavior. Additionally, chemomechanical effects on shale may also be investigated in the context of the simple micromechanical model proposed here. Sickle \emph{et al.} \cite{2024_Sickle_et_al_Mica} found that such effects in muscovite mica can be explained via the model’s single parameter $\epsilon$, representing the dynamic weakening behavior of the material. 

\subsection{\label{subsec:microseismicity}Microseismicity}
\begin{figure*}
\includegraphics[]{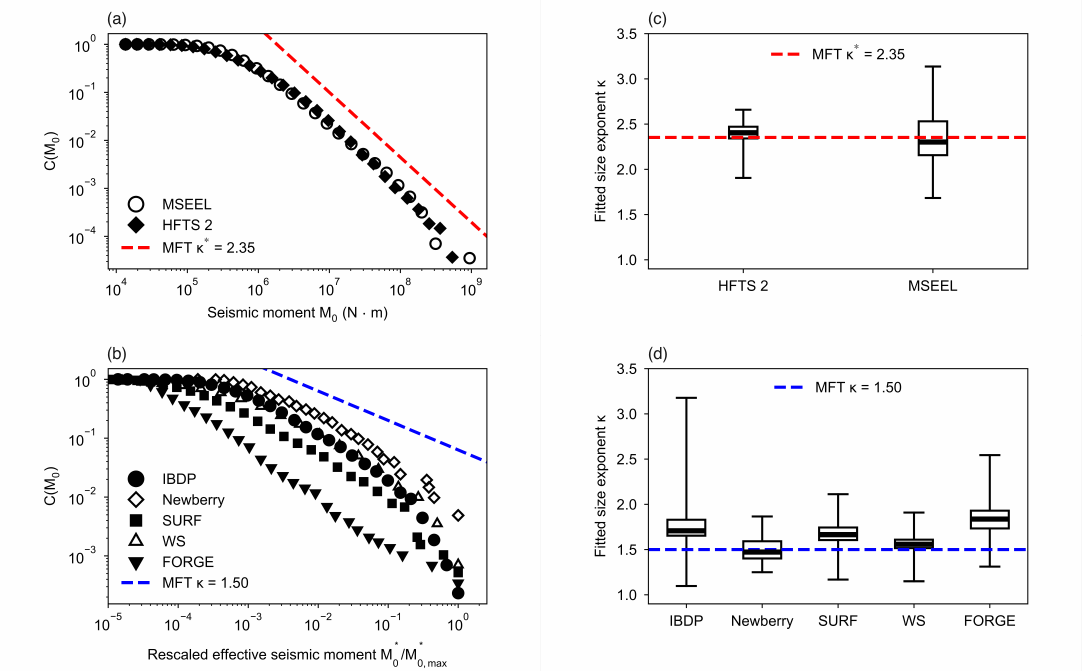}
\caption[]{
Microseismic event distributions. Scalar seismic moment CCDFs for all microseismic catalogs used in this article.
(a) CCDF of microseismic events extracted from operations targeting sedimentary formations. Reference slope shows the MFT prediction for the seismic moment CCDF exponent $\kappa^*$ when stress- and length-scale effects are accounted for (see Discussion section for more details). 
(b) CCDF of microseismic events extracted from operations targeting non-sedimentary formations. Reference slope shows the MFT prediction for the seismic moment CCDF exponent $\kappa$. 
(c) Box-and-whisker plot showing fitted exponents for distributions found in panel (a). Red horizontal dashed line shows the average of the two median exponents. See Methods for a description of the fitting method used to extract these exponents and their 95\% confidence intervals. 
(d) Box-and-whisker plot showing fitted exponents for distributions found in panel (b). Blue horizontal dashed line shows the average of the five median exponents.
}
\label{fig:figure_2}
\end{figure*}

In the case of non-sedimentary formations, the seismic moment distributions shown in Fig.~\hyperref[fig:figure_2]{\ref*{fig:figure_2}(b)} exhibit consistent power-law scaling with the MFT-predicted exponent $\kappa=1.50$ (equivalent to $b=0.75$, see Eq. (\ref{eq:b_to_kappa})) contained within each CCDF’s fitted exponent confidence intervals (Fig.~\hyperref[fig:figure_2]{\ref*{fig:figure_2}(d)}). This $\kappa$ value is remarkably close to the frequently reported $\kappa\approx1.66 \left(b=1.00\right)$ across numerous EGS projects \cite{Baisch_Voros_Weidler_Wyborn_2009, 2012_Evans_et_al, 2016_Cladouhos_et_al} and catalogs of larger-scale seismicity \cite{2010_Kagan}, indicating that a universal physical mechanism describing small- and large-scale loading dynamics in the subsurface is well justified. For layered sedimentary formations, the seismic moment distributions shown in Fig.~\hyperref[fig:figure_2]{\ref*{fig:figure_2}(a)} also exhibit power-law scaling, albeit with a higher typical exponent of $\kappa=2.35 \left( b\approx2.00 \right)$. As in the case of non-sedimentary formations, this $\kappa$ value is consistent with those reported in the literature but remains well above that of both the mean-field theory value and the nanoindentation experiments. We argue below that this $\kappa$ value is fully consistent with mean-field theory when the layered structure of sedimentary formations is accounted for. While the spread in both fitted exponent values is notable, the results are consistent with the model predictions within error bars.

Before addressing the increase in the $\kappa$ value, we refer to the work of Kagan \cite{2010_Kagan} for insight into why $b$-values in any catalog may be increased. He proposes the existence of several systematic and random errors that contribute to inflated exponents, the most relevant of which being the following: 
\begin{enumerate}
    \item Over-estimating “corner” or “threshold” moment $\left( M_{\text{t}} \right)$ values (the value at which the power-law seismic moment scaling gives way to a rapidly decaying cutoff function). Fitting a power law with the $M_{\text{t}}$ value located within the rounded cutoff region will increase the $b$-value. The MLE method overestimates $b$ in this case as well \cite{2009_Clauset_et_al}.
    \item Combining events occurring in different tectonic settings that naturally possess different corner moments. While the effects of stress conditions and focal mechanisms are discussed by Schorlemmer \emph{et al.} \cite{2005_Schorlemmer_et_al}, finite length scale effects can also play a role: faults with larger areas naturally permit larger maximum seismic moments, in turn giving rise to relatively large cutoffs. 
    \item Summing the moment tensors of earthquake subevents with different focal mechanisms leads to smaller tensor sums and thus larger exponents.
\end{enumerate}
Possible cause (1) is unlikely to impact all $b$-value calculations, but it may still contribute to a significant fraction of them. Each microseismic catalog exhibits a unique distribution shape, preventing the consistent estimation of both the magnitude of completeness $M_{\text{c}}$ and the corner moment $M_{\text{t}}$. Moreover, the choice of distribution to be used in the fitting procedure (i.e., plain or tapered G-R) is highly variable along with the researchers’ estimations of $M_{\text{c}}$ and $M_{\text{t}}$. Possible causes (2) and (3), however, are likely contributors to the inflated $b$-values measured during nearly all operations that yield microseismicity. It is well known that the small shear slips induced during these operations occur on various natural faulting planes located near the injection site \cite{2018_Eaton}, and as previously mentioned, maximum seismic moment depends in part on the size of these planes. Given the relative abundance of smaller faulting planes, smaller events are generally overrepresented in microseismic catalogs, hence the increased $b$-values. Previous work which determined lower maximum physical $b$-values \cite{1999_Olsson, 2014_Hallo_et_al, 2015_Eaton_Maghsoudi}, near 1.50, do not take these possible causes into account. 

We now turn our attention to the exponent inflation seen in microseismic catalogs extracted from layered sedimentary formations. While some theoretical work on fracture has yielded larger exponents that may be consistent with these very large $b$-values \cite{2018_Baro_et_al, 2001_Pradhan_Chakrabarti}, we do not discuss it here due to the consensus that shear slip rather than fracture is the dominant mechanism of microseismicity. As mentioned earlier, layered sedimentary formations yield significantly higher $b$-values than those yielded by more homogeneous formations. The layered structure permits natural opportunities for shear planes of various sizes and orientations to be pressurized during and after fluid injection; as discussed before, catalogs composed of microseisms occurring in different tectonic settings will overrepresent small events. Previous nanoindentation studies on shales \cite{2021_Wang_et_al} have shown that mechanical properties such as elastic moduli, hardness, and toughness depend on bedding planes’ orientations relative to the indent direction. Local regions of sedimentary rocks may then exhibit spatiotemporal variance in their mechanical properties as hydraulic fractures propagate and cause the surrounding stress field to change.

\begin{figure*}
\includegraphics[]{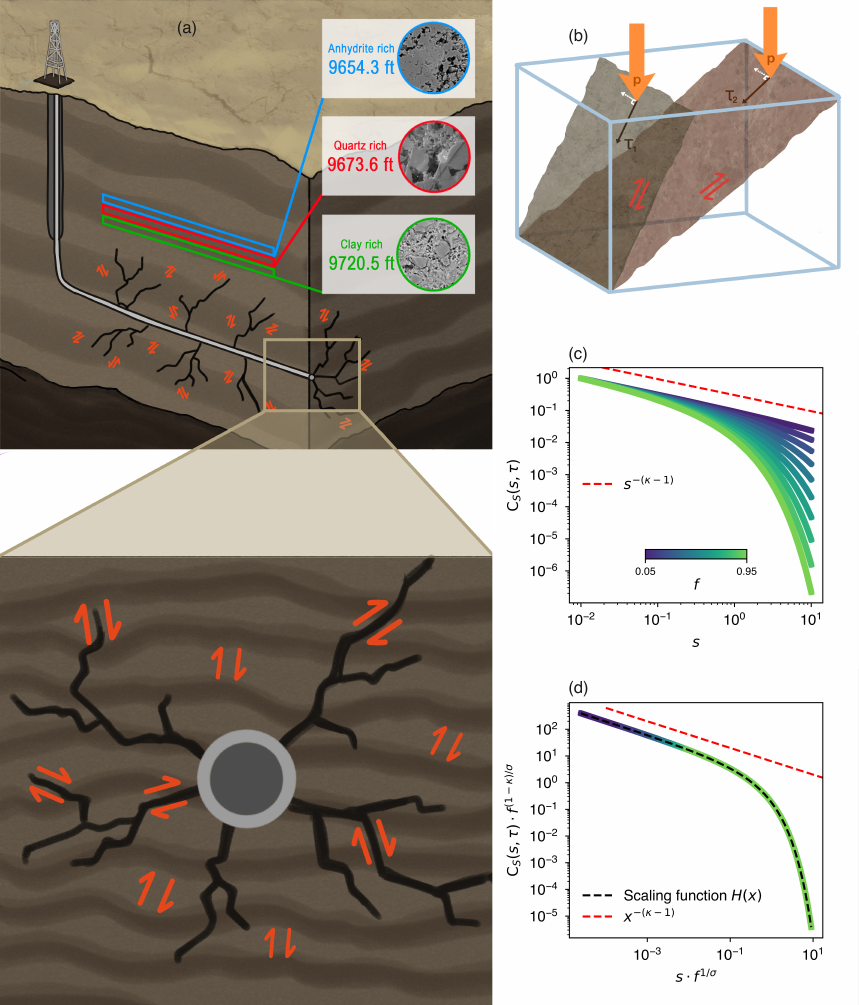}
\caption[]{
Local stress and length scale effects on microseismicity. 
(a) Example hydraulic fracturing schematic. High-pressure fluid injection forces fractures to grow away from the wellbore, inducing slips in the surrounding rock. SEM images and extraction depths of the Cane Creek shale samples used in this study included. 
(b) Two shear planes at different shear stresses $\tau_2 > \tau_1$ under an identical vertical force $p$. 
(c) CCDF form of Eq. (\ref{eq:size_pdf}) referenced in the main text evaluated at various reduced stresses $f=(\tau_{\text{c}}-\tau)/\tau_{\text{c}}$.
(d) Scaling collapse of the curves in panel (c). MFT predicts the universal scaling function $H(x)$.
}
\label{fig:figure_3}
\end{figure*}

In contrast to EGS where faults induced to slip are likely operating near their critical stresses, HF operations induce new tensile fractures that exert varying stresses (ranging from near-zero to near-critical) on faults near the wellbore due to the highly layered nature of the target formations. Examples of different stress conditions imposed on faults can be seen in Fig.~\hyperref[fig:figure_3]{\ref*{fig:figure_3}(b)}. At a fixed stress level $\tau$ and for a fixed linear system size $L$, the probability distribution function $P_S(s,\tau,L)$ of slip sizes $s$ is predicted by the model to scale as 
\begin{equation}
\label{eq:size_pdf}
P_S(s,\tau,L)\sim s^{-\kappa} F\left( s(\tau_{\text{c}}-\tau)^{\frac{1}{\sigma}} \right),
\end{equation}
where $(\tau_{\text{c}}-\tau)^{-\frac{1}{\sigma}} \sim s_{\text{max}} \sim L^{-d_{\text{f}}}$ is the largest slip size with $d_{\text{f}}=2.00$ being the fractal dimension, $\kappa=1.50$ the mean-field scaling exponent, and $F(x,y)$ a rapidly decaying universal scaling function whose scaling form is predicted by the theory for large $L$ \cite{1999_Perkovic_et_al, Friedman_THESIS}. The CCDF form of Eq. (\ref{eq:size_pdf}) in the limit of infinite system size, $C_S(s,\tau)$, can be found in Fig.~\hyperref[fig:figure_3]{\ref*{fig:figure_3}(c)} for various reduced stresses $f=(\tau_{\text{c}}-\tau)/\tau_{\text{c}}$. (Note that for $\tau \rightarrow \tau_{\text{c}}$, the expected power-law behavior of $C_S(s,\tau)$ is recovered). Fig.~\hyperref[fig:figure_3]{\ref*{fig:figure_3}(d)} shows these CCDFs collapsed onto the universal scaling function $H(x)=x^{1-\kappa} G(x)$ predicted by the theory.

Assuming a uniform stress distribution (see Fig.~\hyperref[fig:figure_3]{\ref*{fig:figure_3}} and Supplemental Information), integrating up to the maximum stress $\tau_{\text{max}}$ yields 
\begin{equation}
\label{eq:stress_integrated_size_pdf}
P_S^{\tau_{\text{int}}}(s,L) \sim s^{-(\kappa+\sigma)} \mathcal{F}_\tau\left( s(\tau_{\text{c}}-\tau_{\text{max}})^{\frac{1}{\sigma}},sL^{-d_{\text{f}}} \right),
\end{equation}
where $\sigma=0.50$ in MFT and $\mathcal{F}_\tau(x,y)$ is another universal scaling function (see Supplemental Information for more detailed derivations). Furthermore, based on the results of Bailey \cite{2015_Bailey} which showed that stratigraphic layer thicknesses $h$ are power-law distributed as $P(h) \sim h^{-\alpha}$ with a typical exponent of $\alpha=1.70$, we integrate over linear fault dimensions $L$ up to some maximum scale $L_{\text{max}}$ to yield a final distribution of slip sizes. We first assume a distribution of vertical layer heights $h$ that act as slip-limiting boundaries, then allow for slip planes oriented in all directions to occur in any layer. The resulting marginal probability distribution of slip sizes is
\begin{equation}
\label{eq:stress_and_length_integrated_size_pdf}
P_S^{\tau_{\text{int}},L_{\text{int}}}(s) \sim s^{-\kappa^*} \mathcal{F}_\tau\left( s(\tau_{\text{c}}-\tau_{\text{max}})^{\frac{1}{\sigma}},sL_{\text{max}}^{-d_{\text{f}}} \right),
\end{equation}
where $\kappa^*=\kappa+\sigma+(\alpha-1)/d_{\text{f}}$ and $\alpha$ is the power-law exponent for the layer height distribution. Assuming a value of $\alpha=1.70$, this yields $\kappa^* = 2.35$ (or $b\approx2.00$), in complete agreement with the values found here and in the literature. Given that $\kappa^*$ depends on the power-law exponent $\alpha$, variations in the distribution of sedimentary layer heights from one formation to another may lead to variations in the corresponding $b$-value. This critical exponent, if estimated properly from a sufficiently large catalog, may then yield information about the local distribution of mechanical bed heights. Using the relations $\kappa^*-1=2b/3$ and $\kappa^*=\kappa+\sigma+(\alpha-1)/d_{\text{f}}$, a common $b$-value range of $[1.5,3.0]$ corresponds to an $\alpha$ range of $[1.0,3.0]$. 

Finally, we give a simple heuristic argument for why measuring a $b$-value of 4.00 $(\kappa=4.00)$ or greater is highly unlikely for lack of sufficient amounts of data. Assume that a catalog of seismic moments is distributed according to a power law with exponent $\kappa=4.00$. A catalog that contains only values that perfectly scale over exactly one order of magnitude as $C_S(s) \sim s^{-(\kappa-1)}$, say from size $S$ to $10S$, would then have 1,000 events given that $C_S(10s)=(10s)^{-(\kappa-1)}=(10s)^{-3}=0.001s^{-3}=0.001C_S(s)$. A catalog that scales over two orders of magnitude (common in the literature) would contain 1 million events, too many for even long-term monitoring studies. Our simplifying assumption of this catalog having only perfectly scaling events is also unlikely: generally, magnitudes of completeness and corner magnitudes are well above and below the minimum and maximum events in a catalog, respectively, further increasing the difficulty of fitting to a distribution governed by a power law with exponent $b \ge 4.00$.

\subsection{\label{subsec:future_work}Future Work}
Future nanoindentation experiments will enhance the current work by incorporating higher time resolution measurements and more indents to thoroughly resolve avalanche dynamics in layered samples that experience stress at different angles with respect to layer orientation. The model predicts many scaling laws for which the current experimental resolution is insufficient, including maximum velocity distributions and size- and duration-averaged avalanche shapes. Such average shapes are strict tests of agreement with the theory: whole functions allow us to test for dynamical agreement with model predictions. These quantities could also be extracted from microseismic catalogs that include time-resolved seismic moment and/or relative stress conditions for each earthquake.

\section{\label{sec:methods}Methods}

\subsection{\label{subsec:nanoindentation}Nanoindentation experiments}

Indentation experiments were conducted using a TI 950 TriboIndenter system coupled with an xSol thermal stage at the University of Utah NanoFab facility. By pressing a sharp diamond indenter probe into the surface of a sample under the controlled load, mechanical properties of the material at the micro- to nanoscale were obtained. Prior to indentation, the shale samples were polished to achieve a smooth surface, using sandpapers followed by diamond lapping films down to a $0.05 ~ \mu \text{m}$ grit size. This preparation ensured a smooth surface and the accuracy of the indentation tests. We evaluated the mineralogical distribution in shale using scanning electron microscopy with Energy Dispersive Spectrometry (SEM-EDS) analysis in combination with the box-counting method. Representative elementary area size was determined as $600 ~ \mu \text{m} \times 600 ~ \mu \text{m}$ for the following indentation tests. Each indentation test consisted of three stages: loading, holding, and unloading. The load was increased at a constant rate of 0.3 mN/s for 10 seconds, held at a maximum of 3 mN for 5 seconds, and then unloaded at the same rate. To capture the spatial heterogeneity, 100 indents (10 $\times$ 10 grid) were performed for each sample at each testing temperature: \tempone, \temptwo, \tempthree, \tempfour. A spacing of $60 ~ \mu \text{m}$ between adjacent indents was maintained to avoid interaction effects between neighboring points.

\subsection{\label{subsec:popin_detection}Pop-in detection}

Pop-ins were detected during indentation as a tip displacement event where the slope of the displacement time series remained below the median slope value in a window around each point. We used a sliding window of length 200 timesteps (equivalent to 10\% of the loading period). Detection begins after a load threshold of 100 $\mu$N is crossed due to initial rapid tip displacement into the samples. The effects of both the sliding window length and the threshold load are discussed further in the SI. 

\subsection{Microseismic event extraction from observational data}

All microseismic events used here for analysis were extracted as moment magnitudes or scalar seismic moments directly from the data files. Seismic waveform analysis was performed by the data providers prior to cataloging. 

\subsection{Probability distribution fitting}

The maximum-likelihood method used here for determining best-fit probability distribution parameters is based on the work of Deluca and Corral \cite{2013_Deluca_Corral}; a highly detailed explanation of our specific Python implementation of this method can be found in \cite{Sickle_THESIS}, appendix A. In brief, we seek three quantities for each distribution: (1) the exponent characterizing the distribution of events and (2) the minimum and (3) maximum values between which the power-law probability distribution best describes the data. A power-law distribution is a probability distribution over some random variable $X$ with the following asymptotic form:
\begin{equation}
P_X(x) \sim x^{-\alpha},
\end{equation}
where $\alpha$ is the exponent characterizing the distribution. In real systems, such a probability distribution cannot exactly describe the random variable $X$, given that there must be both a lower and an upper limit on the values $X$ can take. Thus, the distribution is generally defined over some finite range $X \in [x_{\text{min}},x_{\text{max}}]$ and takes the normalized form 
\begin{equation}
\label{eq:normalized_power_law_pdf}
P_X(x) = \frac{\alpha-1}{x_{\text{min}}^{1-\alpha} - x_{\text{max}}^{1-\alpha}} x^{-\alpha}
\end{equation}
in the case of a continuous random variable $X$. In this work we treat both pop-in sizes and microseismic moments as continuous random variables. Modifications to this distribution are necessary in the discrete case \cite{2005_Newman}. 

Note that this approach to extracting distribution parameters is different from those common in geophysics; the most frequently used is that of Aki \cite{1965_Aki}. This approach yields a maximum-likelihood estimate of the $b$-value
\begin{equation}
\hat{b}=\frac{\log{e}}{\langle M\rangle -M_{\text{c}}}
\end{equation}
and its 95\% confidence interval
\begin{equation}
\hat{b}\left( 1-\frac{d_{\epsilon}}{\sqrt{N}} \right) \le \hat{b} \le \hat{b} \left( 1+\frac{d_{\epsilon}}{\sqrt{N}} \right),
\end{equation}
where $d_\epsilon = 1.96$ at the 95\% confidence level, $\langle M \rangle$ denotes the average magnitude for $M \ge M_{\text{c}}$, and $M_{\text{c}}$ is the “magnitude of completeness”, the magnitude level above which it is assumed that, for the region under study, the catalog is complete and there are no missing events. Pardo-Iguzquiza \& Dowd \cite{2024_Pardo-Iguzquiza_Dowd} demonstrate the equivalence of Aki’s method to fitting a continuous, lower-truncated power-law distribution to scalar seismic moments. Efforts have been made to account for the effects of magnitude binning (effectively transforming a continuous random variable into a discrete one), but corrections are insignificant when the bin size is on the order of 0.1 or lower \cite{2003_Marzocchi_Sandri, 1983_Bender}. 

It is well known among researchers studying power-law distributions in data that such fits and their confidence intervals can be heavily biased for several reasons, not the least of which being the choice of bounds for the fit \cite{2009_Clauset_et_al, 2013_Deluca_Corral, 2019_Corral_Gonzalez, 2012_Stumpf_Porter}. Aki’s method may contain biases for at least two reasons: (1) it depends on the choice of $M_{\text{c}}$ and (2) does not employ the fully truncated power-law distribution (Eq. (\ref{eq:normalized_power_law_pdf})) necessary when fitting to data extracted from finite systems with a cutoff. Regarding (1), there is still no consensus on an optimal method for determining the magnitude of completeness $M_{\text{c}}$ \cite{2018_Eaton, 2018_Kozlowska_et_al}. Methods include using the peak of the non-cumulative event histogram (dependent on binning choices), using the point of maximum curvature of the magnitude-frequency distribution, and modeling the catalog for the entire magnitude range (EMR) to then determine the point at which the distribution becomes nonlinear \cite{2019_Zorn_WV_marcellus_shale, 2000_Wiemer_Wyss}. Clauset \emph{et al.} \cite{2009_Clauset_et_al} demonstrate that care must be taken to appropriately account for both $x_{\text{min}}$ and $x_{\text{max}}$, otherwise erroneous exponents and their confidence intervals can be extracted from maximum likelihood estimation. Varying techniques for identifying the power-law scaling range found are thus likely culprits for the highly variable $b$-values found within and between microseismic analyses. 

\section*{Acknowledgements}
The authors would like to thank Isabel Zhou for her work designing the figures and Charles Choens for helpful discussions on shale. E.A.M. would like to thank Jingqiang Tan, Lei Li, Ge Jin, Erich Zorn, Mark Zoback, and Elliot Jagniecki for helpful discussions and guidance on data collection. This work was funded by Sandia National Laboratories Laboratory Directed Research and Development (LDRD) program, project \#233080. Y.W. and P.N. were sponsored by the Department of Energy under DOE Award Number DE-FE0031775, titled “Improving Production in the Emerging Paradox Oil Play.” This article has been authored by employees of National Technology \& Engineering Solutions of Sandia, LLC under Contract No. DE-NA0003525 with the U.S. Department of Energy (DOE). The employees own all right, title and interest in and to the article and is solely responsible for its contents. The United States Government retains and the publisher, by accepting the article for publication, acknowledges that the United States Government retains a non-exclusive, paid-up, irrevocable, world-wide license to publish or reproduce the published form of this article or allow others to do so, for United States Government purposes. The DOE will provide public access to these results of federally sponsored research in accordance with the DOE Public Access Plan \url{https://www.energy.gov/downloads/doe-public-access-plan}. This paper describes objective technical results and analysis. Any subjective views or opinions that might be expressed in the paper do not necessarily represent the views of the U.S. Department of Energy or the United States Government.

\section*{Data Availability}
The seven microseismic catalogs analyzed in this work
are publicly available from their respective data
repositories: the Marcellus Shale Energy and Environment
Laboratory (MSEEL) \cite{MSEEL_Data}, Hydraulic
Fracturing Test Site 2 (HFTS-2) \cite{HFTS2_Data},
Illinois Basin--Decatur Project (IBDP) \cite{IBDP_Data},
Newberry EGS \cite{Newberry_Data}, EGS Collab at the
Sanford Underground Research Facility (SURF)
\cite{EGSCollab_Data}, Utah FORGE \cite{FORGE_Data},
and WHOLESCALE \cite{WHOLESCALE_Data}.
The nanoindentation measurements and processed pop-in data are
available from the authors upon reasonable request. 

\section*{Author Contributions}
P.N., Y.W., A.G.I., and F.W.D. designed the nanoindentation experiments, and Y.W., A.G.I., and F.W.D. performed the nanoindentation experiments. A.G.I. performed the thermogravimetric analysis. E.A.M. and J.J.S. wrote the nanoindentation analysis code. E.A.M. wrote the microseismic analysis code, generated the plots, and wrote the paper. E.A.M. and K.A.D. interpreted the results. K.A.D. provided project supervision. All authors reviewed and edited the manuscript and provided critical feedback.

\section*{Competing interests}
The authors declare no competing interests.

\bibliography{references}

\end{document}